\documentclass[10pt,oneside,letterpaper]{article}
\usepackage{amssymb}
\usepackage{amsmath}
\usepackage[dvips]{graphicx}

\begin{document}

\baselineskip=18pt
\numberwithin{equation}{section}
\allowdisplaybreaks  

\pagestyle{myheadings}

\thispagestyle{empty}

\vspace*{-2cm}
\begin{flushright}
{\tt hep-th/yymmnnn}\\
\end{flushright}

\vspace*{2.5cm}
\begin{center}
 {\LARGE Non Supersymmetric Metastable Vacua in ${\cal N}=2$\\  SYM Softly Broken to ${\cal N}=1$\\}
 \vspace*{1.7cm}
 Georgios Pastras$^1$\\
 \vspace*{1.0cm}
  $^1$ Jefferson Physical Laboratory, Harvard University, Cambridge, MA 02138,
USA\\
 \vspace*{0.8cm}
 {\tt pastras@fas.harvard.edu}
\end{center}
\vspace*{1.5cm}

\noindent

We find non-supersymmetric metastable vacua in four dimensional
${\cal N}=2$ gauge theories softly broken to ${\cal N}=1$ by a
superpotential term. First we study the simplest case, namely the
$SU(2)$ gauge theory without flavors. We study the spectrum and
lifetime of the metastable vacuum and possible embeddings of the
model in UV complete theories. Then we consider larger gauge group
theories with flavors. We show that when we softly break them to
${\cal N}=1$, the potential induced on specific submanifolds of
their moduli space is identical to the potential in lower rank gauge
theories. Then we show that the potential increases when we move
away from this submanifold, allowing us to construct metastable
vacua on them in the theories that can be reduced to the $SU(2)$
case.

\newpage
\setcounter{page}{1}

\section{Introduction}

Dynamical supersymmetry breaking in a metastable vacuum is an
attractive possibility for supersymmetry breaking. Unlike
old-fashioned spontaneous supersymmetry breaking one can consider
candidate theories with supersymmetric vacua elsewhere in field
space. In a beautiful paper \cite{Intriligator:2006dd} showed that
this scenario is realized even in simple ${\cal N}=1$ gauge theories
like SQCD with massive flavors. Since \cite{Intriligator:2006dd}
there has been a lot of activity in the direction of extending the
results in field theory \cite{Kitano:2006wm, Banks:2006ma,
Schmaltz:2006qs, Dine:2006xt, Kitano:2006xg, Murayama:2006yf,
Csaki:2006wi, Intriligator:2007py} and string theory
\cite{Franco:2006es, Ooguri:2006pj, Ooguri:2006bg, Franco:2006ht,
Bena:2006rg, Argurio:2006ny, Aganagic:2006ex, Giveon:2007fk,
Argurio:2007qk, Kawano:2007ru} realizations.

It was already pointed out in \cite{Intriligator:2006dd} that it
might be interesting to study the system of ${\cal N}=2$
supersymmetric theories softly broken to ${\cal N}=1$ by
superpotential terms. ${\cal N}=2$ theories have moduli spaces of
vacua. Unlike ${\cal N}=1$, in ${\cal N}=2$ it is possible to
compute the Kahler metric on the moduli space exactly. If we add a
small superpotential, we can hope that we can still use the exact
Kahler metric. This allows us to compute the scalar potential on the
moduli space exactly and look for local minima that correspond to
metastable non-supersymmetric vacua.

In this paper we study the simplest example, namely pure ${\cal
N}=2$ $SU(2)$ gauge theory, softly broken to ${\cal N}=1$ by a
superpotential for the scalar field. For appropriate selection of
the superpotential a metastable vacuum appears at the origin of the
moduli space. We discuss the spectrum of the theory in this vacuum,
its lifetime and possible embeddings of our model in a UV complete
theory. Then we consider ${\cal N}=2$ theories with gauge groups of
higher ranks and with flavors. We show that on specific submanfolds
of their moduli space the potential is identical with the potential
of lower rank theories. We also show that these submanifolds can be
locally stable allowing us to construct metastable vacua on them as
in $SU(2)$.

While this paper was being prepared for publication another paper appeared
\cite{Ooguri:2007iu}, which has overlap with this work.

\section{Pure ${\cal N}=2$ $SU(2)$ gauge theory}

\subsection{The metric on the moduli space}

The field content of pure ${\cal N}=2$ $SU(2)$ gauge theory consists
of the gauge field $A_\mu$, a complex scalar $\phi$ and fermions, all in the adjoint
representation of the gauge group. The theory has a moduli space of vacua, in
which the gauge group is broken to $U(1)$, that we will refer to as
 the
Coulomb branch. The classical potential for the scalar field $\phi$
in ${\cal N}=2$ SYM without flavors is:
\begin{equation}
V\left( \phi  \right) = \frac{1}
{{g^2 }}Tr\left( {\left[ {\phi ,\phi ^\dag  } \right]} \right)
\end{equation}
Setting the potential to zero gives the semi-classical moduli space of vacua,
characterized by
a complex number multiplying the element of the Cartan subgroup of the gauge
group:
\begin{equation}
\phi  = \frac{1}{2} \left( {\begin{array}{*{20}c}
   a & 0  \\
   0 & { - a}  \\

 \end{array} } \right)
\end{equation}
where $a$ is a complex number. However $a$ is not a gauge invariant
quantity, so we identify the vacua by the complex number
\begin{equation}
u = Tr\phi ^2  = \frac{1} {2}a^2
\end{equation}

Using the powerful constraints of  ${\cal N}=2$ supersymmetry, one
can go beyond the semiclassical analysis and study the full quantum
theory. In the seminal paper \cite{Seiberg:1994rs} Seiberg and
Witten managed to determine exactly the low energy effective theory
on the Coulomb branch. The quantum moduli space turns out to be the
complex $u$-plane with singularities. Classically one expects a
singularity at $u=0$ where the $SU(2)$ gauge symmetry is restored.
It turns out that quantum mechanically the point $u=0$ is smooth and
there is no gauge symmetry enhancement anywhere on the moduli space.
Instead, there are two singularities at $u=\pm 1$ \footnote{More
precisely the two singularities are at the points $u=\pm \Lambda$.In
this paper we are using units where the scale $\Lambda=1$.} where
monopoles and dyons become massless.

The exact Kahler metric on the moduli space was computed in
\cite{Seiberg:1994rs} and can be written in the following form:

\begin{equation}
  ds^2  = g(u) du d\bar u = \operatorname{Im} \left( {\tau \left( u \right)} \right)\left| {\frac{{da \left( u \right)}}
{{du}}} \right|^2 dud\bar u \label{kahlermetric}
\end{equation}
where:

\begin{equation}
\begin{gathered}
  \tau \left( u \right) = \frac{{\frac{{da_D \left( u \right)}}
{{du}}}} {{\frac{{da\left( u \right)}}
{{du}}}} \hfill \\
  a\left( u \right) = \sqrt 2 \sqrt {u + 1} {}_2F_1 \left( { - \frac{1}
{2},\frac{1}
{2};1;\frac{2}
{{u + 1}}} \right) \hfill \\
  a_D \left( u \right) = i\frac{{u - 1}}
{2}{}_2F_1 \left( {\frac{1}
{2},\frac{1}
{2};2;\frac{{1 - u}}
{2}} \right) \hfill \\
\end{gathered}
\end{equation}

\subsection{Soft breaking to ${\cal N}=1$}

We now consider adding a superpotential for the chiral multiplet,
breaking ${\cal N}=2$ down to ${\cal N}=1$. If the superpotential
term is small, we can assume that we can still trust the effective
IR description of the theory. In other words we assume that the
Kahler metric on the moduli space is the same as in usual ${\cal
N}=2$ SW $U(1)$ theory and that the effect of the superpotential is
to induce a superpotential $W(u)$ for the effective IR scalar field
$u$. This superpotential will produce a potential on the moduli
space equal to:

\begin{equation}
V\left( u \right) = g^{ - 1} \left( u \right)\left|W'(u)\right|^2
\label{scalarpotential}
\end{equation}
where the Kahler metric is still given by the above relations
\eqref{kahlermetric}.

\begin{figure}[h]
\begin{center}
\includegraphics[angle=0,width=0.7\textwidth]{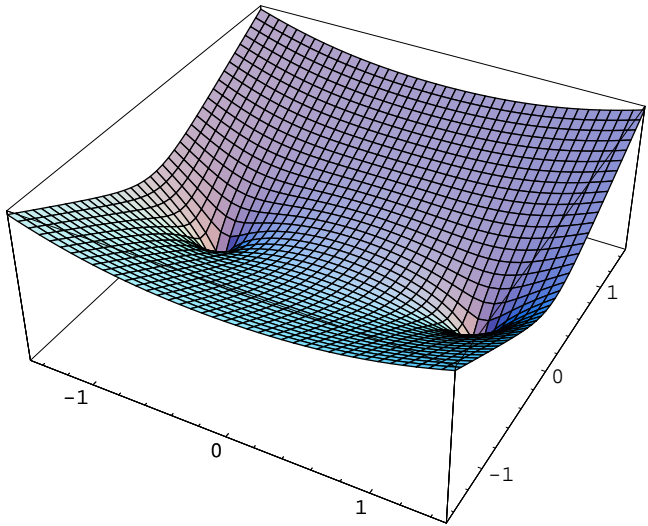}
\end{center}
\caption{The potential due to the quadratic superpotential}
\label{fig:vsw}
\end{figure}

The goal of this paper is to find a superpotential $W(u)$ which, once
combined with the Kahler metric $g(u)$ given by the Seiberg-Witten solution,
will induce a scalar potential \eqref{scalarpotential} with a local minimum
at some point of the moduli space. Of course this minimum must have nonzero
energy if it has to correspond to a non-supersymmetric metastable vacuum.
As noticed in
\cite{Intriligator:2006dd}, the simplest
choice is the superpotential $W\sim Tr\Phi^2$,
in terms of the UV fields, which takes the form $W(u) \sim u$ in terms
of the fields in the IR effective theory:

\begin{equation}
\begin{gathered}
  W = \mu u \hfill \\
  V\left( u \right) = \mu^2 g^{ - 1} \left( u \right) \hfill \\
\end{gathered}
\end{equation}
In this case the potential is equal to the inverse Kahler metric
multiplied by a constant and we can see it plotted in figure \ref{fig:vsw}.

As was pointed out in \cite{Intriligator:2006dd} in this case there are no
metastable vacua in the moduli space. We can see the two usual supersymmetric
vacua for the ${\cal N}=2$ theory broken to ${\cal N}=1$ by a mass
term for the adjoint, which were also described in \cite{Seiberg:1994rs}. They
correspond to the points on the moduli space where non perturbative objects
became massless. We can also see a saddle point between them at the origin
$u=0$.

The next step is to consider a more general superpotential. From the
relation \eqref{scalarpotential} we see that the scalar potential in
the general case is the product of two factors: of the inverse
Kahler metric and of the square of the  derivative of the
superpotential. As we saw above, the inverse Kahler metric has two
global minima at $u=\pm 1$ where it is equal to zero corresponding
to the supersymmetric vacua and a saddle point at $u=0$. We will try
to find a superpotential that will produce a local minimum for the
scalar potential at $u=0$, where the inverse Kahler metric has a
saddle point.

Since the
function $W(u)$ is holomorphic, it is easy to show that $|W'(u)|^2$ cannot
have local minima except for the supersymmetric ones, when $W'(u)=0$. However
$|W'(u)|^2$ can have saddle points. By choosing $W(u)$ appropriately we
can arrange that the saddle point of $|W'(u)|^2$ lies at $u=0$, so that
it coincides with the saddle point of the inverse Kahler metric. It is not
difficult to show that the product of two functions which have a common
saddle point at some $u_0$ will also have a stationary point at $u_0$.
Moreover we can see that depending on the relative magnitudes of the
second partial derivatives of the two functions, it is possible that the
product can have a {\it local minimum} at $u_0$ even if the two factors
only have saddle points at $u_0$.

In our case it turns out that the simplest possibility to consider
is a superpotential of third order in $u$:
\begin{equation}
W = \mu \left( u + \lambda u^3 \right)
\end{equation}

We have set the quadratic term in $u$ to  zero, so that the saddle
point of $|W'(u)|^2$ occurs exactly at the origin $u=0$. To have a
chance to get a metastable vacuum we need the stable and unstable
directions of the saddle point of $|W'(u)|^2$ to be related to the
stable and unstable directions of the saddle point of $g^{-1}$ in
such a way that the product of the two functions $V(u) = g^{-1}
|W'(u)|^2 $ has a stationary point with all directions stable.
Otherwise we would again get a saddle point. This occurs if the
coefficient of the third order term is positive.

\begin{figure}[h]
\begin{center}
\includegraphics[angle=0,width=0.7\textwidth]{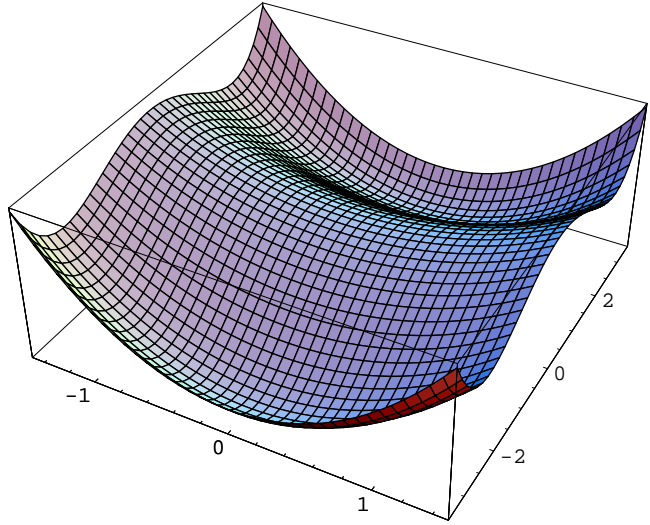}
\end{center}
\caption{The potential by a third order superpotential with flat
Kahler metric} \label{fig:vw}
\end{figure}

In figure \ref{fig:vw} we can see the plot of $|W'(u)|^2$, which
indeed has a saddle point as we wanted. So now the potential is
going to be the product of the two graphs above. The second
derivatives of $|W'(u)|^2$ at the saddle point increase as we
increase $\lambda$. So we expect if $\lambda$ is too small, the
saddle point to look like the first graph, while if $\lambda$ is too
large to look like the second graph, as we can understand by the
following relation.

\begin{equation}
\begin{gathered}
  \left. {\frac{{d^2 V}}
{{d\operatorname{Re} \left( u \right)^2 }}} \right|_0  = \mu ^2 \left( {\left. {12\lambda g^{ - 1} } \right|_0  + \left. {\frac{{d^2 g^{ - 1} }}
{{d\operatorname{Re} \left( u \right)^2 }}} \right|_0 } \right) \hfill \\
  \left. {\frac{{d^2 V}}
{{d\operatorname{Im} \left( u \right)^2 }}} \right|_0  = \mu ^2 \left( {\left. { - 12\lambda g^{ - 1} } \right|_0  + \left. {\frac{{d^2 g^{ - 1} }}
{{d\operatorname{Im} \left( u \right)^2 }}} \right|_0 } \right) \hfill \\
\end{gathered}
\label{eq:der}
\end{equation}

\begin{figure}[h]
\begin{center}
\includegraphics[angle=0,width=0.7\textwidth]{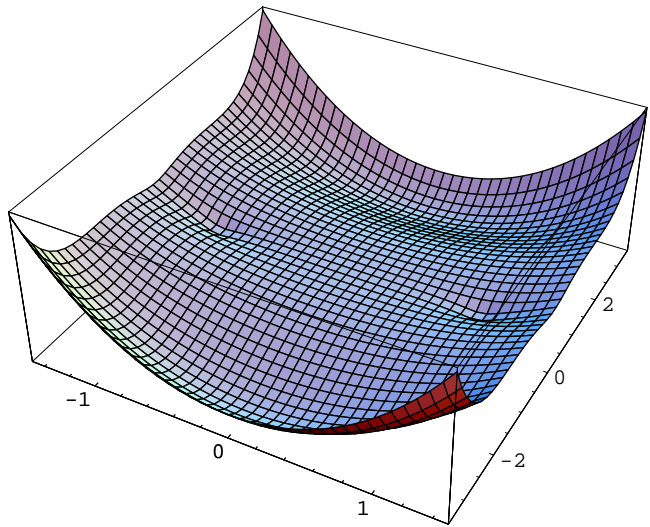}
\end{center}
\caption{The full potential, a metastable vacuum exists  at the
origin} \label{fig:pot}
\end{figure}

\begin{figure}[ph]
\begin{center}
\includegraphics[angle=0,width=0.7\textwidth]{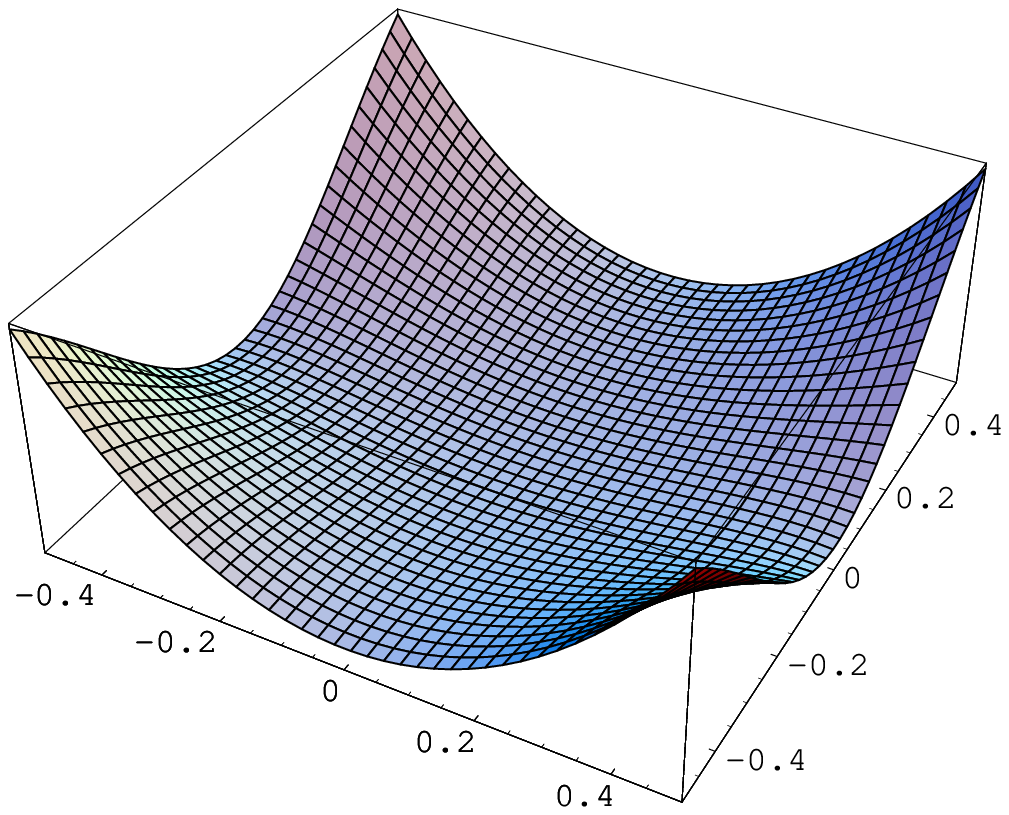}
\end{center}
\caption{Close-up on the x-saddle point} \label{fig:cl1}
\begin{center}
\includegraphics[angle=0,width=0.7\textwidth]{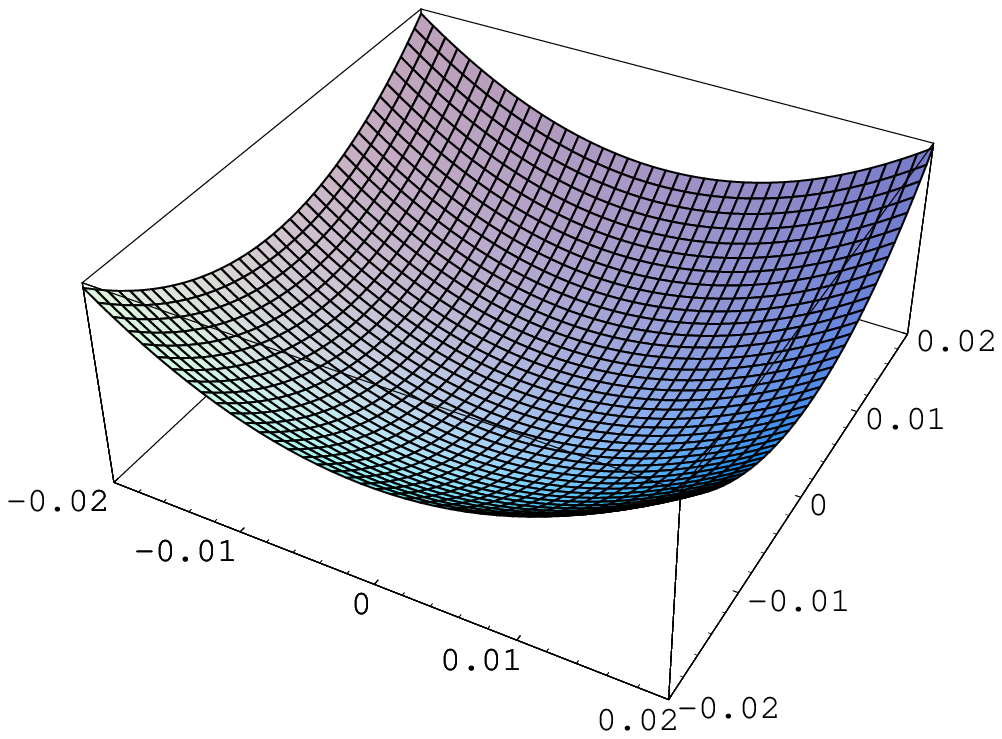}
\end{center}
\caption{A closer close-up on the x-saddle point} \label{fig:cl2}
\end{figure}

Fortunately $ \left|  {\frac{{d^2 g^{ - 1} }} {{d\operatorname{Re}
\left( u \right)^2 }}} \right| < \left| {\frac{{d^2 g^{ - 1} }}
{{d\operatorname{Im} \left( u \right)^2 }}} \right|$ at the origin,
so actually there is a range of $\lambda$ for which the origin
becomes a local minimum. Using properties of the hypergeometric
functions we find:

\begin{equation}
  \lambda _ -   < \lambda  < \lambda _ +   \hfill
\end{equation}
where

\begin{equation}
\lambda _ \pm   = \frac{1} {{24}}\left[ {1 \pm \left( {\frac{{\Gamma
\left( {\frac{3} {4}} \right)}} {{2\Gamma \left( {\frac{5} {4}}
\right)}}} \right)^4} \right]
\end{equation}
For example, for superpotential equal to:

\begin{equation}
W = 0.01\left( u + \frac{1}{24} u^3 \right)
\end{equation}
we get the picture of figure \ref{fig:pot}. If one zooms at the
x-saddle point (figures \ref{fig:cl1} and \ref{fig:cl2}) sees the
meta-stable vacuum.

Let's make a few comments on this potential. There are four
supersymmetric vacua. Two are the Seiberg Witten ones at $u=\pm 1$.
The other two are the ones induced by the superpotential, they are
the zero's of $\left| \frac {\partial W}{\partial u} \right|^2$.
Their position is $u=\pm i \frac{1}{\sqrt{3 \lambda}}$. The
metastable vacuum lies at $u=0$, and has four possible decays
towards the four supersymmetric vacua.

Another thing is that we can make the ${\cal N}=2$ breaking
superpotential as small we like by making $\mu$ small. Changing
$\mu$ results just in multiplication of the potential with a
constant thus not changing the picture we saw above. So the
assumption we made that there is no significant change induced to
the Kahler metric by this superpotential can be satisfied.

The picture of the potential, as we move in this parameter region
changes as follows.

\begin{figure}[ph]
\begin{center}
\includegraphics[angle=0,width=0.7\textwidth]{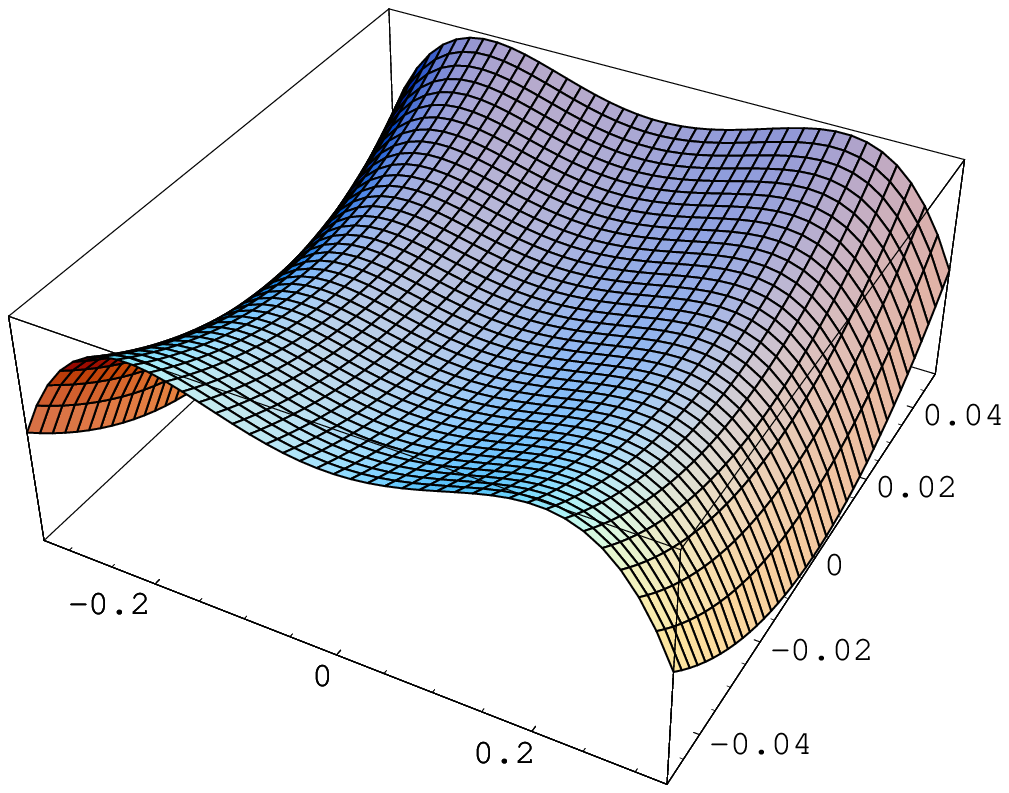}
\end{center}
\caption{The area of the metastable vacuum for $\lambda$ close to
$\lambda_-$} \label{fig:lm}
\begin{center}
\includegraphics[angle=0,width=0.7\textwidth]{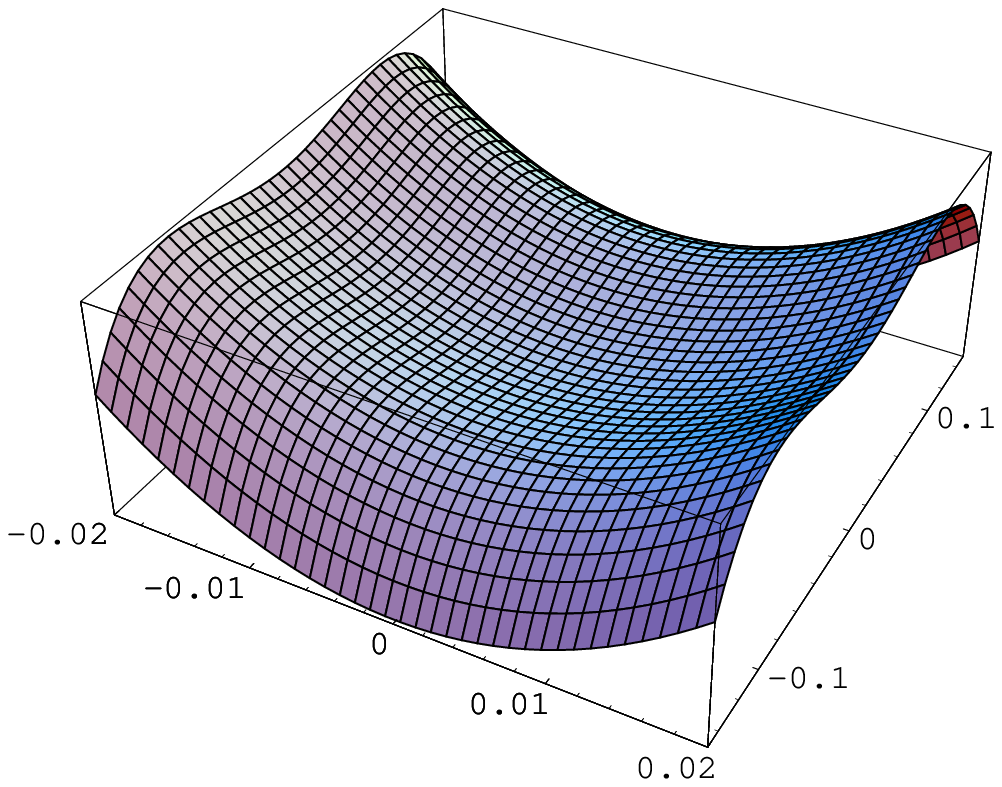}
\end{center}
\caption{The area of the metastable vacuum for $\lambda$ close to
$\lambda_+$} \label{fig:lp}
\end{figure}

For $\lambda$ close to $\lambda_-$ as seen in figure \ref{fig:lm}
the metastable vacuum minimum is elongated along the imaginary axis,
looking more possible to decay towards the Seiberg-Witten vacua.

For $\lambda$ close to $\lambda_+$ as seen in figure \ref{fig:lp}
the metastable vacuum minimum is elongated along the real axis,
looking more possible to decay towards the superpotential vacua.

\subsection{Lifetime of the metastable vacuum}

In order to estimate the lifetime of the metastable vacuum, we use
the triangular approximation \cite{Duncan:1992ai}. The thin wall
approximation \cite{Coleman:1977py} is not a good approximation in
our case, as the ratio of the barrier width and barrier height is
not small enough.

Let's say the position and potential at the true vacuum, the false
vacuum and the peak of the potential between them are $\phi _ -$,
$\phi _ + $, $\phi_T$ and $V _ -$, $V _ + $, $V_T$ respectively.
Then we define

\begin{equation}
\begin{gathered}
  \Delta \phi _ \pm   = \left| {\phi _T  - \phi _ \pm  } \right| \hfill \\
  \Delta V_ \pm   = V_T  - V_ \pm   \hfill \\
  \lambda _ \pm   = \frac{{\Delta V_ \pm  }}
{{\Delta \phi _ \pm  }} \hfill \\
  c = \frac{{\lambda _ -  }}
{{\lambda _ +  }} \hfill \\
\end{gathered}
\end{equation}
and the decay rate is given by:

\begin{equation}
\frac{\Gamma }
{V} \sim Ae^{ - B}
\end{equation}
B is given by,

\begin{equation}
B = \frac{{32\pi ^2 }}
{3}\frac{{1 + c}}
{{\left( {\sqrt {1 + c}  - 1} \right)^4 }}\left( {\frac{{\Delta \phi _ +  ^4 }}
{{\Delta V_ +  }}} \right)
\end{equation}
Applying the above in our case for both possible decays we get:

\begin{equation}
\begin{gathered}
  c_{SW}  = \frac{{V_{T,SW} }}
{{V_{T,SW}  - V_0 }}\frac{{\phi _{T,SW} }}
{{1 - \phi _{T,SW} }} \hfill \\
  c_W  = \frac{{V_{T,W} }}
{{V_{T,W}  - V_0 }}\frac{{\phi _{T,W} }}
{{\frac{1}
{{\sqrt {3\lambda } }} - \phi _{T,W} }} \hfill \\
\end{gathered}
\end{equation}

\begin{figure}[ph]
\begin{center}
\includegraphics[angle=0,width=0.7\textwidth]{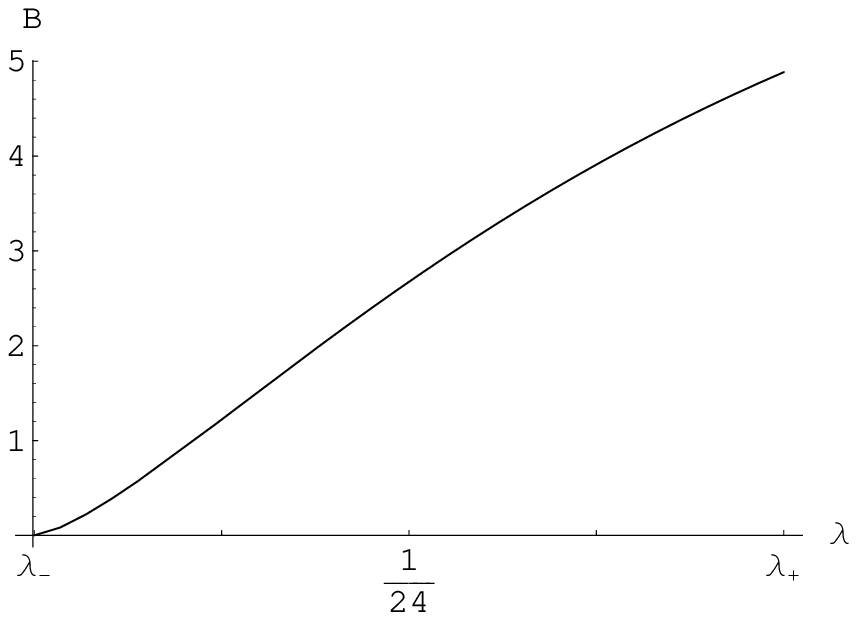}
\end{center}
\caption{$B$ factor for the decay towards the SW vacuum}
\label{fig:bsw}
\begin{center}
\includegraphics[angle=0,width=0.7\textwidth]{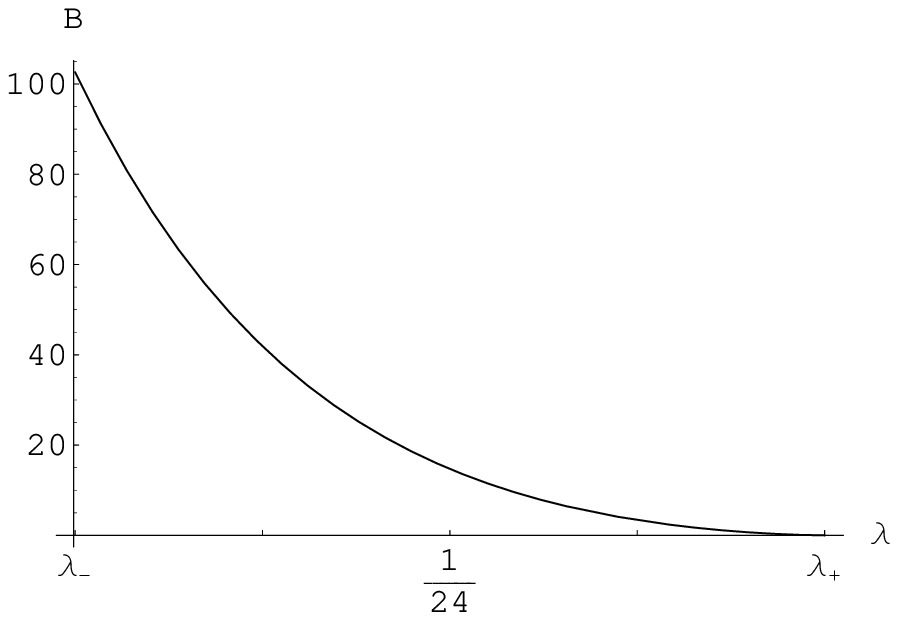}
\end{center}
\caption{$B$ factor for the decay towards the superpotential vacuum}
\label{fig:bw}
\end{figure}

The numerical results for the $B$ factor as function of the parameter
$\lambda$ for $\mu =1$ in this approximation are shown in figures
\ref{fig:bsw} and \ref{fig:bw}.

We observe that generally decay towards the SW vacua is favorable.
This is because the superpotential vacua are more distant. We also
observe that for $\lambda=\lambda_-$ the field flows directly
towards the SW vacuum, and for $\lambda=\lambda_+$ towards the
superpotential vacuum as expected. There is actually a value for
$\lambda$ around $0.0457$ where the two decays are equally
favorable.

Above we found the decay rate in the triangular approximation for
$\mu=1$. If one assumes an arbitrary $\mu$ what happens is:

\begin{equation}
\begin{gathered}
  \phi_ \pm  \to \phi_ \pm  \hfill \\
  V_ \pm  \to \mu ^2 V_ \pm  \hfill \\
  c \to c \hfill \\
  B \to \mu ^{ - 2} B \hfill \\
\end{gathered}
\end{equation}
meaning that we can make the metastable vacuum as long lived as we
want by making $\mu$ as small as necessary. Notice that our analysis
is reliable exactly for small $\mu$ as noticed above.

\subsection{Spectrum at the metastable vacuum}

Obviously as there is a $U(1)$ gauge symmetry remaining everywhere
in the moduli space, the gauge boson remains massless. The gaugino
is also massless.

The mass of the fermion partner of $u$ is given by the second
derivative of the superpotential. However as we have not added any
second order terms in the superpotential the second derivative at
the origin is zero. That means that the fermion partner of $u$
remains also massless.

Last we can calculate the scalar masses directly from the second
derivatives of the scalar potential. We are careful to divide with
the inverse Kahler metric as our fields are not canonical. From
equations \ref{eq:der} we find:

\begin{equation}
\begin{gathered}
   m^2_{\phi\operatorname{Re}}  = 12\mu ^2
  {\left( \lambda -\lambda_- \right)  }   \hfill \\
   m^2_{\phi\operatorname{Im}}  = 12\mu ^2
  {\left( \lambda_+ -\lambda \right)  }   \hfill \\
\end{gathered}
\end{equation}
The supertrace equals

\begin{equation}
\sum {m_B ^2  - \sum {m_F ^2  = 12\mu ^2 \left( {\lambda _ +   -
\lambda _ -  } \right) } } = \left( {\frac{{\Gamma \left( {\frac{3}
{4}} \right)}} {{2\Gamma \left( {\frac{5} {4}} \right)}}} \right)^4
\mu ^2
\end{equation}
and does not depend on $\lambda$ as expected (because the third
derivatives of the superpotential give off-diagonal contributions to
the boson mass matrix). The supertrace is non zero as the Kahler
metric has non vanishing curvature.

As we have two massless fermions it is a fair question which is the
goldstino. The gaugino does not interfere to the process of
supersymmetry breaking and this is actually the reason it remains
massless, or better its mass does not split with the mass of the
gauge boson. The actual goldstino is the fermion partner of the
scalar $u$

\subsection{Origin of the non-renormalizable superpotential}

We know that the non-trivial Kahler metric we use occurs in the low
energy limit of a theory well defined in the UV, specifically the
${\cal N}=2$ $SU(2)$ SYM without flavors. We would like to know if
it is also possible to get also the non-renormalizable
superpotential as low energy effective superpotential of a UV
complete theory. We have to be careful though, so we don't alter the
${\cal N}=2$ theory in such a way that the SW Kahler metric
calculation is not reliable anymore.

It is actually simple to get the appropriate term by adding in the
${\cal N}=2$ SYM a gauge singlet massive hypermultiplet $( \tilde
M$, $M)$. In order to get something non trivial for the adjoint we
need to couple it with the heavy hypermultiplet. The only possible
gauge invariant and renormalizable term is $MTr\Phi ^2$. We also add
a third order superpotential for the other chiral multiplet, so:

\begin{equation}
W = aMTr\Phi ^2  + b\tilde M^3  + mM\tilde M
\end{equation}

For the reliability of the Kahler metric calculation we need the
mass term has to be large so, the extra hypermultiplet can be
integrated out at high enough energies. The other two parameters
have to be small, so they don't alter the structure of the ${\cal
N}=2$ at low energies.

In order to integrate out the massive hypermultiplet, we need to
find the equations of motion of the massive fields, as they occur
from the superpotential, and substitute the massive hypermultiplet
in the superpotential, using these equations. We find:

\begin{equation}
W_{eff} =  - \frac{{a^3 b}} {{m^3 }}\left( {Tr\Phi ^2 } \right)^3
\end{equation}
Thus at low energies we get the required superpotential. As the
coefficient of this term in our analysis was $\mu \lambda$, and we
could make $\mu$ as small as we like, that means that we can make
$a$, $b$ and $m^{-1}$ as small as needed for the validity of the
calculation.

\section{Higher rank groups}

\label{sec:hrg}

\subsection{$SU(3)$ ${\cal N}=2$ SYM with flavor}

In the case of $SU(3)$ the moduli space is a two dimensional complex
manifold parametrized by the two complex numbers $u$, $v$. The
metric on the moduli space can be computed from the curve $C$
\cite{Klemm:1994qs, Argyres:1994xh}:

\begin{equation}{y^2 = (x^3 -u x -v )^2 -1}\end{equation}
where we have set the strong coupling scale $\Lambda=1$.

When we add flavor the curve takes the form \cite{Hanany:1995na,
Krichever:1996ut}:

\begin{equation}
{y^2 = (x^3 - u x - v)^2 -\prod_{i=1}^{N_f} (x+m_i)}
\label{eq:curve}
\end{equation}
where $m_i$ are the masses of the hypermultiplets.

Using Riemann bilinear identities we can write the Kahler metric
elements as:

\begin{equation}
\begin{gathered}
g_{u,\bar{u}} = \int_C {\omega_u\wedge \bar{\omega_u}},
\quad g_{v,\bar{v}} = \int_C {\omega_v \wedge \bar{\omega_v}} \hfill \\
g_{u,\bar{v}} = \int_C {\omega_u \wedge \bar{\omega_v}},
\quad g_{v,\bar{u}} = \int_C {\omega_v \wedge \bar{\omega_u}} \hfill \\
\end{gathered}
\end{equation}
where

\begin{equation}{\omega_v = {1\over y} dx,\qquad \omega_u = {x\over y} dx}
\end{equation}
Then the potential equals

\begin{equation}{V = g^{u,\bar{u}} |W_u|^2 + g^{v,\bar{v}} |W_v|^2 +
2 \operatorname{Re} (g^{u,\bar{v}} W_u
W_v^*)}\label{pot}\end{equation}
where

\begin{equation}
W_u=\frac{\partial W}{\partial u}, \quad W_v=\frac{\partial
W}{\partial v}
\end{equation}

This is a function of the variables $u$, $v$, the parameters $m_i$,
and the derivatives of the superpotential $W_u$, $W_v$. For any
given superpotential and values of the parameters we have to find
local minima in terms of the variables $u$, $v$.

Unfortunately as we can see from \eqref{pot} we have to invert the
Kahler metric to write down the potential and this complicates
things a little bit. Inverting the Kahler metric, we can write the
potential in the form:

\begin{equation}{V = {1\over g_{u\bar{u}} g_{v\bar{v}} -
|g_{u\bar{v}}|^2}\times (g_{v\bar{v}} |W_u|^2 + g_{u\bar{u}}|W_v|^2
+ 2 \operatorname{Re}(g_{u\bar{v}}W_u\bar{W}_v))}
\end{equation}
The potential has the form:

\begin{equation}{V = {h_1\over h_2}}\end{equation}
where

\begin{equation}
\begin{gathered}
h_1 = (g_{v\bar{v}} |W_u|^2 + g_{u\bar{u}}|W_v|^2 + 2
\operatorname{Re}(g_{u\bar{v}}W_u\bar{W}_v))\\
h_2 = g_{u\bar{u}} g_{v\bar{v}} - |g_{u\bar{v}}|^2
\end{gathered}
\end{equation}

The coefficients of the Kahler metric that we want to compute to
determine the effective potential have the form:

\begin{equation}
g_{a\bar{b}} (z) = \int dx d\bar{x} \frac{g(x)} {|P(x)|}
\end{equation}
where $g(x)$ is either $|x|^2$ or $x$ or constant and $P(x)$ is the
right hand side of \ref{eq:curve}

We are interested in calculating the above in locations of the
moduli space where $P(x)$ has a double root. We show in appendix
\ref{sec:app1} although the metric elements go to zero as the above
integral diverges, the potential is actually finite and has the
form:

\begin{equation}
{V_{eff} ={h_1\over h_2} = {|W_u|^2 + |W_v|^2 |r|^2 + 2
\operatorname{Re} (r W_u W_v^*) \over \int dx d\bar{x}{1\over
|P'(x)|}}} \label{eq:pot}
\end{equation}
where $r$ is the double root and

\begin{equation}{P(x) = P'(x) (x-r)^2}\end{equation}
which is simplified, but more importantly, it is finite.

So to minimize the potential on the singular submanifold we have to
maximize:

\begin{equation}{\int dx d\bar{x}{1\over |P'(x)|}}\end{equation}
where remember that $P'(x)$ is the factorized curve. Except for this
expression being simpler to calculate numerically than the initial
one, it is possible, as the factored polynomial is a lower rank one,
that the potential on the singular submanifold matches the potential
of a lower rank theory. We will show later that this actually
happens.

\subsection{$SU(3)$ with 2 massless flavors}

In $SU(3)$ with even number of flavors or with no flavors, the sixth
order polynomial factorizes to two third order polynomials making
analysis much simpler. We do this analytically for general even
number of massive flavors in appendix \ref{sec:app2}. Here we use
the results only for two massless flavors.

In $SU(3)$ with 2 massless flavors we have:

\begin{equation}
P = \left( {x^3  - ux - v} \right)^2  -  {x } ^{2}
\end{equation}

We observe that at $v=0$ the polynomial has a double root equal to
zero. According to the previous analysis the potential on this
surface equals:

\begin{equation}
{V_{eff} ={|W_u|^2 \over \int dx d\bar{x}{1\over |P'(x)|}}}
\end{equation}
where

\begin{equation}
P' = \left( {x^2  - u} \right)^2  - 1
\end{equation}

As the double root in this case is $r=0$ the potential on the
submanifold, as seen from equation \ref{eq:pot}, the potential
depends only on $W_u$.

So the potential is identical with the $SU(2)$ theory without
flavors, which we analyzed in previous section. So it suffices to
show that at $u=0$ the potential increases as we move away from the
singular submanifold, to show that we can again construct a
metastable vacuum, using the exact same softly breaking
superpotential.

We saw that $W_v$ does not make any difference in the potential on
the submanifold. However it is interesting to see what effects such
a term has close to the submanifold. For simplicity let's assume a
superpotential term:

\begin{equation}
W_v  = \kappa v
\end{equation}

We show in appendix \ref{sec:app3} that the potential induced by
this term for small $v$ is approximately:

\begin{equation}
V_v \sim \frac{{\left( {1 - \left| 2 v \right|} \right)}} {{\log
\left( {\left| 2 v \right|} \right)}} \end{equation}

\begin{figure}[h]
\begin{center}
\includegraphics[angle=0,width=0.7\textwidth]{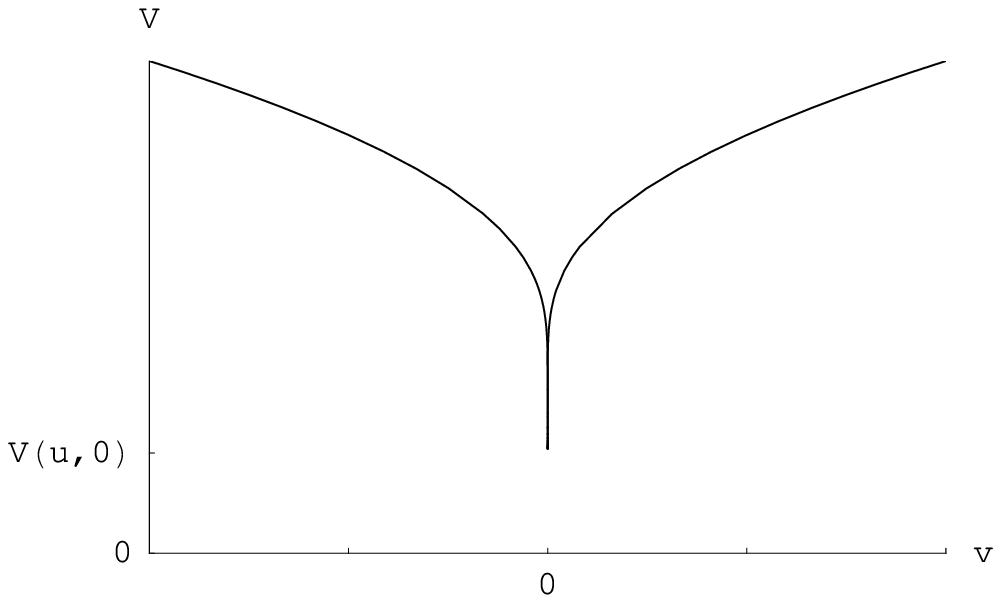}
\end{center}
\caption{Potential in a perpendicular to the singular submanifold
direction} \label{fig:stab}
\end{figure}

This potential strongly constrains the field in the singular
submanifold, as one can see in figure \ref{fig:stab}. So once we
turn on the $W_v$ term, the $v=0$ plane becomes locally stable.

Moreover this result combined with the result of the previous
section, means that ${\cal N}=2$ $SU(3)$ SYM with two massless
flavors, and softly breaking to ${\cal N}=1$ superpotential

\begin{equation}
W = \mu \left( {u + \lambda u^3 } \right) + \kappa v
\end{equation}
has a metastable vacuum at $u=v=0$, for the same range of $\lambda$
as in the pure $SU(2)$ case.

\subsection{Comments on higher rank groups}

It seems clear, that the results analyzed in previous two sections
and the appendices can be generalized to higher rank groups. $SU(N)$
theory with $N_f$ massless flavors can be reduced to $SU(N-1)$ with
$N_f-2$ massless flavors on the subspace where the highest order
moduli is zero. Moreover this subspace can be locally stabilized by
turning on a superpotential term linear in this moduli.

This means that we can construct a metastable vacuum in the origin
of the moduli space of all $SU(N)$ theories with $2N-4$ massless
flavors.

Constructions of metastable vacua in theories with large flavor
symmetries can have interesting phenomenological applications in
particular in building models of direct gauge mediation. (See for
example \cite{Giudice:1998bp} for references)

We can also show that $SU(3)$ with two massive flavor reduces to
$SU(2)$ on the $v=0$ submanifold with the only additional change of
a shift of the moduli $u$ by $\frac{3}{4} m^2 $ (See appendix
\ref{sec:app2}. This means that we are again able to build the same
metastable vacuum if we simply shift $u$ in the appropriate
superpotential too. However a more careful analysis for the local
stability of the submanifold is needed.

Higher rank groups without flavors are studied in
\cite{Ooguri:2007iu}.

\section*{Acknowledgments}

I would like to thank N.Arkani-Hamed and D.Shih for useful
discussions. I would like to especially thank K.Papadodimas for
turning my attention to the subject, and for introducing me to the
subject of SW techniques. I would also like to thank J.Marsano and
K.Papadodimas for the long collaboration that gave lots of the
results of the third section. This work was partially supported by
DOE grant \# DE-FG02-91ER40654.

\appendix

\section{Potential on singular submanifolds}
\label{sec:app1}

\subsection{The potential in $SU(3)$ ${\cal N}=2$ SYM with flavor}

We saw in section \ref{sec:hrg} that the potential has the form:

\begin{equation}{V = {h_1\over h_2}}\end{equation}
where

\begin{equation}
\begin{gathered}
h_1 = (g_{v\bar{v}} |W_u|^2 + g_{u\bar{u}}|W_v|^2 + 2
\operatorname{Re}(g_{u\bar{v}}W_u\bar{W}_v))\\
h_2 = g_{u\bar{u}} g_{v\bar{v}} - |g_{u\bar{v}}|^2
\end{gathered}
\end{equation}

The function $h_1$ depends on $u$, $v$, $m_i$ and also $W_u$, $W_v$,
while $h_2$ is independent of $W_u$, $W_v$.

The function $h_1$ can be written in the form:

\begin{equation}{h_1 = \int dx d\bar{x} { |W_u|^2 + |x|^2 |W_v|^2 +
2 \operatorname{Re}(x W_u W_v^*)\over y \bar{y}}}\end{equation} or
\begin{equation}{h_1 = \int dx d\bar{x} {|W_u + x W_v|^2 \over
    |(x^3 -u x -v)^2 - \prod_{i=1}^{N_f}(x+m_i)|}}\end{equation}

To sum up the moduli space is parametrized by two complex variables
$u$, $v$. The effective scalar potential on the moduli space is
equal to:

\begin{equation}{ V_{eff} = {h_1(u,v)\over h_2(u,v)}}\end{equation}
where the functions $h_1(u,v)$, $h_2(u,v)$ also depend on $W_u$,
$W_v$ and the masses of the quarks and can be written in terms of
integrals over the x-plane as:

\begin{equation}{h_1 = \int dx d\bar{x} {|W_u + x W_v|^2 \over
    |(x^3 -u x -v)^2 - \prod_{i=1}^{N_f}(x+m_i)|}}\end{equation}

\begin{multline}
h_2 =\left( \int dx d\bar{x} {|x|^2 \over
|(x^3 -u x -v)^2 - \prod_{i=1}^{N_f}(x+m_i)|} \right)\times \\
\times \left(  \int dx d\bar{x} {1 \over |(x^3 -u x -v)^2 -
\prod_{i=1}^{N_f}(x+m_i)|}\right) \\
- \left| \int dx d\bar{x} {x \over |(x^3 -u x -v)^2 -
\prod_{i=1}^{N_f}(x+m_i)|}\right|^2
\end{multline}

\subsection{Toy Model}

We are interested in calculating the potential at submanifolds of
the moduli space where the polynomial has a double root. Obviously
the above integrals diverge on such submanifolds. However we will
show that the inverse Kahler metric is finite. We can use the
following way of regularizing the integral:

\begin{equation}{k(z) = \int { f(x) \over |x||x-z|}}\end{equation}
with the function $f(x)$ smooth near zero and falling off fast
enough at infinity. Again the integral diverges logarithmically as
$z\to 0$. We add and subtract the function $f(0) \over |x| (|x|+1)
(|x| + |z|)$ and we have:

\begin{multline}
k(z) = \int\left( {f(x) \over |x||x-z|} -{f(0) \over |x| (|x|+1)
(|x| + |z|)} \right)\\ + \int {f(0) \over |x| (|x|+1) (|x| + |z|)}
\end{multline}
this is of the form:

\begin{equation}{k(z) = h(z) + g(z)}\end{equation}
where we can compute the second function exactly:

\begin{equation}{g(z) = 2\pi f(0) {\log(z) \over 1-z}}\end{equation}
and the first function is finite for all values of $z$. So the only
divergence as $z\rightarrow 0$ comes from $g(z)$ and we can write:

\begin{equation}{k(z) = - 2 \pi f(0) \log(z) + \int \left ({f(x) \over |x|^2} -{f(0) \over |x|^2 (|x|+1)}  \right) + {\cal O}(z)  }\end{equation}
Of course if the double root is not at zero but at an other point
$x=r$ we have the obvious generalization:

\begin{multline}
\int { f(x) \over |x-r||x-r-z|} \\
= - 2 \pi f(r) \log(z) + \int \left ({f(x) \over |x-r|^2} -{f(r)
\over |x-r|^2 (|x-r|+1)} \right) + {\cal O}(z)
\end{multline}

\subsection{Application to our system}

The SW curve for our system (SU(3) with flavor) has the form:

\begin{equation}{y^2 = P(X)}\end{equation}
where $P(x) = \prod_i (x-r_i)$ is some polynomial in $x$, whose
coefficients depend on the moduli space coordinates $u$, $v$.

Let's consider a point on the moduli space $(u_0,v_0)$ where two and
only two roots coincide, let's say to the value $r$.

This means that at that point the polynomial $P$ factorizes to:

\begin{equation}{P(x) = P'(x) (x-r)^2}\end{equation}
where $P'$ is a polynomial of degree two less than $P$ and with all
roots distinct. Now let's move away from the singular point
$(u_0,v_0)$ in such a way that the distance between the
ex-coincident roots is $z$, where by $z$ we denote the absolute
value of the distance.

The coefficients of the Kahler metric that we want to compute to
determine the effective potential have the form:

\begin{equation}{g_{a\bar{b}} (z) \sim \int dx d\bar{x} {g(x) \over |P(x)|}=\int dx d\bar{x} {f(x) \over |(x-r)(x-r-z)|}}\end{equation}
where $g(x)$ is either $|x|^2$ or $x$ or constant and $f(x)$
contains both $g(x)$ and the polynomial $P'(x)$.

As we can see this integral diverges logarithmically as
$z\rightarrow 0$.

We want to write the diverging integrals in the form:

\begin{equation}{g_{a\bar{b}}(z) = -a \log(z) + b + {\cal O}(z)}\end{equation}
and determine the constants $a$ and $b$ in terms of $f(x)$:

Using our previous trick we write:

\begin{multline}
g_{a\bar{b}}(z) = - 2 \pi f(r) \log(z) \\
+ \int \left ({f(x) \over |x-r|^2} -{f(r) \over |x-r|^2 (|x-r|+1)}
\right) + {\cal O}(z)
\end{multline}

Let's do it for all the coefficients of the Kahler metric:

\begin{equation}{g_{u\bar{u}} = \int {|x|^2 \over |P(x)| } = \int {|x|^2 \over |P_z'(x)||x-r||x-r-z|}}\end{equation}
where the subscript $z$ in $P'$ means the $P'$ maybe depends slowly
on $z$ but in the limit $z\rightarrow 0$ this dependence is
irrelevant.

As before we have:

\begin{multline}
g_{u\bar{u}} = -2 \pi {|r|^2 \over |P'(r)|} \log(z) \\
+ \int \left({|x|^2\over |x-r|^2 |P'(x)|} - {|r|^2 \over |P'(r)|
|x-r|^2 (|x-r|+1)} \right) +{\cal O}(z)
\end{multline}
Similarly we find:

\begin{multline}
g_{v\bar{v}} = -2 \pi {1\over |P'(r)|} \log(z) \\
+ \int \left({1\over |x-r|^2 |P'(x)|} - {1 \over |P'(r)| |x-r|^2
(|x-r|+1)} \right) +{\cal O}(z)
\end{multline}
and finally:

\begin{multline}
g_{u\bar{v}} = -2 \pi {r \over |P'(r)|} \log(z) \\
+ \int \left({x\over |x-r|^2 |P'(x)|} - {r \over |P'(r)| |x-r|^2
(|x-r|+1)} \right) +{\cal O}(z)
\end{multline}

In the denominator of the effective potential we have the
combination \linebreak $h_2= g_{u\bar{u}} g_{v \bar{v}} -
|g_{u\bar{v}}|^2$. The terms that go like $(\log z)^2$ cancel, and
we collect all terms proportional to $\log (z)$ to find:

\begin{multline}
h_2 = -2 \pi {1\over |P'(r)|} \log(z) \times \\
\times \left[ \left(\int {|x|^2\over |x-r|^2
|P'(x)|} - {|r|^2 \over |P'(r)| |x-r|^2 (|x-r|+1)} \right) \right.\\
+ |r|^2 \int \left({1\over |x-r|^2 |P'(x)|} - {1 \over |P'(r)|
|x-r|^2 (|x-r|+1)} \right) \\
\left. - 2 \operatorname{Re} \left(r^* \int \left({x\over |x-r|^2
|P'(x)|} - {r \over |P'(r)| |x-r|^2 (|x-r|+1)}\right)\right)\right]
\end{multline}

Now all three integrals are finite, so we can combine them and
simplify the integrand and we find:

\begin{equation}{h_2 = - 2\pi {1\over |P'(r)|}  \log(z) \int dx d\bar{x}{1\over |P'(x)|}+{\cal O}(1)}\end{equation}

The numerator is $ {h_1 = |W_u|^2 g_{v\bar{v}}+|W_v|^2
g_{u\bar{u}}+2 \operatorname{Re} |W_u W_v^*|^2 g_{v\bar{u}}}$ We
keep only the $\log$ diverging part so we find:

\begin{equation}{h_1 = -2\pi {1\over |P'(r)|} \log(z) \left( |W_u|^2 + |W_v|^2 |r|^2 + 2 \operatorname{Re} (r W_u W_v^*)\right)+{\cal O}(1)}\end{equation}
Combining our results we find that the effective potential on the
submanifold where two roots coincide has the simple form:

\begin{equation}
{V_{eff} ={h_1\over h_2} = {|W_u|^2 + |W_v|^2 |r|^2 + 2
\operatorname{Re} (r W_u W_v^*) \over \int dx d\bar{x}{1\over
|P'(x)|}}}
\end{equation}
which is simplified, but more importantly, it is finite.

\section{$SU(3)$ with $N_f=2k$}
\label{sec:app2}

In $SU(3)$ with even number of flavors or with no flavors, the sixth
order polynomial factorizes to two third order polynomials making
analysis much simpler. For this factorization to be possible it
suffices that hypermultiplets come in pairs of equal mass. However
in the following analysis we will assume that they are all equal for
simplicity.

In $SU(3)$ with $2k$ flavors (including $k=0$) we have:

\begin{equation}
P_{2k} = \left( {x^3  - ux - v} \right)^2  - \left( {x + m}
\right)^{2k}
\end{equation}
So we can write:

\begin{equation}
P_{2k} = P_{2k+} P_{2k-}
\end{equation}
where

\begin{equation}
  P_{2k\pm}  = \left( x^3  - ux - v \right) \pm \left( {x + m} \right)^k  \hfill \\
\end{equation}

One possibility is that two roots of $P_-$ or two roots of $P_+$
coincide. However here we will concentrate in a much simpler to
analyze case the case where one root of $P_-$ coincides with one
root of $P_+$. We are mainly interested in this case as we will be
able to show in next section, that the submanifold of the moduli
space where this happens is energetically preferred at least
locally.

Obviously one root of $P_-$ cannot coincide with one root of $P_+$
in the pure $SU(3)$ case. In the other two cases the only
possibility is that the common root is $x=-m$. This means that the
singular submanifold is:

\begin{equation}
v = um - m^3
\end{equation}
On the submanifold the polynomials can be written as:

\begin{equation}
  P_\pm  = \left( {x + m} \right)\left[ {\left( {x^2  - mx + m^2  - u} \right) \pm 1} \right] \hfill \\
\end{equation}
for the two flavors case, and:

\begin{equation}
 P_\pm  = \left( {x + m} \right)\left[ {\left( {x^2  - mx + m^2  - u} \right) \pm \left( {x + m} \right)} \right] \hfill \\
\end{equation}
for the four flavor case.

This means that:

\begin{equation}
P' = \left( {x^2  - mx + m^2  - u} \right)^2  - 1
\end{equation}
for two flavors and

\begin{equation}
P' = \left( {x^2  - mx + m^2  - u} \right)^2  - \left( {x + m}
\right)^2
\end{equation}
for four flavors.

As $x$ is about to be integrated on the whole complex plane, and the
only x-dependence of the integrand is in the polynomial, we can
always shift it by a constant. So we can eliminate the linear terms
in the first parenthesis shifting $x$ by $\frac{m}{2}$. We get:

\begin{equation}
P' = \left( {x^2  + \frac{3} {4}m^2  - u} \right)^2  - 1
\end{equation}
and

\begin{equation}
P' = \left( {x^2  + \frac{3} {4}m^2  - u} \right)^2  - \left( {x +
\frac{3} {2}m} \right)^2
\end{equation}

The last forms are well known. The first is the polynomial for the
pure $SU(2)$ where we have shifted $u$ by $\frac{3}{4}m^2$ and the
second is the $SU(2)$ with $u$ shifted by the same amount and two
flavors of mass $\frac{3}{2}m$.

Specifically for $m=0$ the first case simplifies to the most known

\begin{equation}
P' = \left( {x^2  - u} \right)^2  - 1
\end{equation}
and the singular submanifold in this case is simply $v=0$. As the
double root in this case is $x=0$ the potential on the submanifold,
as seen from equation \ref{eq:pot}, the potential depends only on
the part of the superpotential that depends only on $u$.

\begin{equation}{V_{eff} ={|W_u|^2 \over \int dx d\bar{x}{1\over |P'(x)|}}}\end{equation}

\section{Stability of the singular submanifold}
\label{sec:app3}

In previous appendix we showed that we can have an easier expression
of the potential in a submanifold of the $v=0$ moduli space of
$SU(3)$ gauge theory with $2k$ flavors. Now let's see if the system
has a preference to lay in this singular submanifold. We saw in
previous section that the v-part of the superpotential does not make
any difference in the potential on the submanifold. However it is
interesting to see what effects such a term has close to the
submanifold. For simplicity let's assume a superpotential term:

\begin{equation}
W_v  = \kappa v
\end{equation}

When one moves a little away from the $v=0$ plane, the roots of the
two polynomials slightly move away from zero. For small $v$ the
roots are also small, and one can neglect the cubic term giving us:

\begin{equation}
r_ \pm   \simeq \frac{v} {{u \mp 1}} \Rightarrow z = \left| r_ +   -
r_ - \right| \simeq 2\left| v \right|
\end{equation}

So the difference between the two roots depends linearly on $v$.
This approximation does not hold only close to $u = \pm 1$, but
these are the SUSY SW vacua, so they are stable anyway.

We saw in previous sections that in general wherever two roots of
the polynomial coincide the Kahler metric vanishes logarithmically.
We also saw that the inverse Kahler metric element does not have to
vanish as the divergent parts cancel at the inversion process.

We remind that the logarithmically divergent parts of Kahler metric
elements were:

\begin{equation}
\begin{gathered}
  g_{u\bar u}  =  - 2\pi \frac{{\left| r \right|}}
{{\left| {P'\left( r \right)} \right|}}\frac{{\log \left( z
\right)}}
{{1 - z}} +{\cal O}(1) \hfill \\
  g_{v\bar v}  =  - 2\pi \frac{1}
{{\left| {P'\left( r \right)} \right|}}\frac{{\log \left( z
\right)}}
{{1 - z}} +{\cal O}(1) \hfill \\
  g_{u\bar v}  =  - 2\pi \frac{r}
{{\left| {P'\left( r \right)} \right|}}\frac{{\log \left( z
\right)}}
{{1 - z}} +{\cal O}(1) \hfill \\
\end{gathered}
\end{equation}
where $z$ is the magnitude of the difference of the two approaching
roots, and $r$ is the double root. However in our case the double
root is equal to zero, meaning that there are no divergences in
$g_{u\bar u}$ and $g_{u\bar v}$. That means that if we go a little
bit away of the $v=0$ submanifold, the contribution of the new
superpotential term will be:

\begin{equation}
V_v  = \frac{{g_{u\bar u} \left| {\kappa } \right|^2 }} {{g_{u\bar
u} g_{v\bar v}  - \left| {g_{u\bar v} } \right|}} \sim \frac{{\left(
{1 - \left| 2 v \right|} \right)}} {{\log \left( {\left| 2 v
\right|} \right)}} \end{equation}
plus something similar coming from
the $g_{u\bar v}$ term

\end{document}